\documentclass[
utf8,%
aps,%
12pt,%
final,%
notitlepage,%
oneside,%
onecolumn,%
nobibnotes,%
nofootinbib,%
superscriptaddress,%
noshowpacs,%
]%
{revtex4}

\usepackage[utf8]{inputenc}
\usepackage[T2A]{fontenc}
\usepackage[russian,english]{babel}

\begin{document}

\title{
  On hydrostatic approximation by R.I.~Nigmatulin\\ and L.F.~Richardson's equation.      
}

\author{\firstname{I.N.}~\surname{Sibgatullin}}
\email{sibgat@ocean.ru}
\affiliation{%
Shirshov Institute of Oceanology of Russian Academy of Sciences
}%

\begin{abstract}
  \noindent
The theorem given 
in ''Equations of Hydro-and Thermodynamics of the Atmosphere when Inertial Forces Are Small in Comparison with Gravity'' (2018) is wrong,
since the solutions of the system of Navier-Stokes equations do not converge to the solutions of the system of hydrostatic approximation equations,
when the vertical acceleration approaches zero.
The main consequence is that 
the scales given in the paper are not suitable for 
application of hydrostatic approximation%
. 
The correct asymptotics 
should be given by
the traditional hydrostatic parameter H/L, where H and L are the vertical and horizontal scales of motion.
Also scale analysis of the L.F.~Richardson's equation for vertical velocity in hydrostatic approximation is not correct.
\end{abstract}

\maketitle

\noindent
Keywords: {\it
\noindent
hydrostatic approximation, 
quasistatic approximation, 
Richardson's equation,
synoptic scales,
microscale meteorology,
mesoscale meteorology,
force of inertia
}

\selectlanguage{English}
\begin{sloppypar}

\section{Incorrectness of the theorem formulated 
in~\cite{2018NigmatulinEquations} ,
and scales of applicability of hydrostatic (quasistatic) approximation}

\vspace{-3mm}
\noindent
Traditional asymptotic analysis of the hydrostatic approximation for different geophysical flows
is based on the smallness of the ratio $\varepsilon=H/L$ of the vertical to the horizontal scales of motion, 
which is often introduced as the \emph{hydrostatic parameter} 
(see, f.e., R.~Zeytounian, \cite{1991ZeytounianMeteorologicalFluidDynamics}, eq. 3.9, 19.2).
The author of~\cite{2018NigmatulinEquations} proposed a different approach,
which is 
based only on the smallness of the vertical acceleration normalized by the gravity acceleration,
as the parameter of applicability of the hydrostatic approximation, and formulated it as a theorem.
Smallness of the amplitude of oscillations, which corresponds to the smallness of the acceleration at a fixed frequency, usually can imply only linearization of the equations. 
But until now, no one could formulate a theorem on elimination of the short-wave motions and application of the hydrostatic approximation (which is a long-wave approximation according to the traditional asymptotic analysis) based only on the smallness of amplitude or the vertical acceleration.

The convergence of the solutions of the Navier-Stokes equations to the solutions of the system of hydrostatic approximation equations 
\emph{when the vertical acceleration approaches zero does not exist}, 
since this is directly contradicted by the 
\emph{finite vertical acceleration} of the nontrivial solutions of the hydrostatic 
  approximation equations%
, in which \emph{the equation with the vertical acceleration is replaced by the hydrostatic balance.}
In~\cite{2018NigmatulinEquations} the author also did not give a proof of the existence of such a convergence or transition to the limit. 
Therefore, in view of the indicated contradiction, the theorem \emph{formulated in~\cite{2018NigmatulinEquations} is not true}.


The author applies his own or unusual terminology (``inertialess vertical velocity'', ``climatic scales''), which is usually not necessary. However, here is the source of confusion and the reason for the erroneous conclusions, which are discussed below. 
If a new theorem or method is introduced, the words and definitions must be very clear, and allow to follow logical conclusions.
In the author's opinion, the vertical force of inertia is ``neglected'' in the (quasi) hydrostatic approximation. This statement is incorrect, since the vertical acceleration of any nontrivial solutions of the \emph{system of equations of hydrostatic approximation} is not equal to zero, while the magnitude of the force of inertia is identically equal to the product of mass and acceleration.

In the language of dynamics, that is in terms of the forces acting on the fluid, the following statement would be correct:
the \emph{restoring forces} (with respect to the state of hydrostatic equilibrium) in the vertical direction are eliminated in the equations of the hydrostatic approximation, as compared to the full equations
(in other words, the equation for the momentum in the vertical direction is replaced by the hydrostatic balance). 
But this in no way can be said about the force of inertia in the vertical direction, since due to the action of the forces in the horizontal direction (which are present in the equations for the horizontal momentum, primarily the horizontal pressure gradient) and the equation of continuity, which in this case plays the role of a kinematic constraint, there is an accelerated motion in the vertical direction%
\footnote{
  The force of inertia and the restoring force in vertical direction coincide for the case of the full Navier-Stokes model, since the vertical acceleration is explicitly present in the equation for vertical momentum. In the hydrostatic approximation, the restoring force in vertical direction is eliminated, and the equation for the vertical momentum is replaced by the hydrostatic balance. However, the vertical component of the force of inertia in the hydrostatic approximation is not zero for any nontrivial solutions, so the statement that it is neglected is wrong.
}.

Force of inertia of an atmospheric particle is defined by the sum of the applied forces. Scale of its vertical component in absence of acoustic perturbations is defined by the maximum difference between gravity force and buoyancy, i.e. by deviation of the density in the atmospheric flow under consideration, multiplied by {\bf g}. Resulting {\sl reduced gravity} acceleration can be by three orders 
smaller than {\bf g}, and nevertheless it directly generates important atmospheric flows: convection and internal waves. 
Background 
stratification and thermodynamic properties such as adiabaticity and phase changes have to be taken into account and naturally lead to the concepts of \emph{potential and virtual temperatures}~\cite{Charney1990}.

This is why neglecting the vertical inertia (see again the footnote$^1$) upon comparing it directly to the gravity (as in \cite{2018NigmatulinEquations}) gives \emph{also} absolutely wrong estimation of the spatial and temporal scales, for which hydrostatic approximation can be applied. Hydrostatic framework has no explicit equation for the change of vertical momentum, which is replaced by hydrostatic balance, so the pressure is equal to the weight of the atmospheric column. Instead, vertical velocity (and so the change of vertical momentum) can be calculated with the help of Richardson's equation~\cite{1922RichardsonWPNP,2006LynchEmergence,1974KasaharaVariousCoordinates}, following from the continuity and thermodynamic equations. For the horizontal scales from only 1 km, time from 100 s, vertical velocities below than 1 m/s, given in the paper~\cite{2018NigmatulinEquations}, hydrostatic approximation is incorrect even if the smaller scale motions are filtered (I.~Kibel, A.~Obukhov etc.), because for vertical force of inertia being two or three orders of magnitude less than gravity, hydrostatic approximation can not describe important types of atmospheric flows and gives absolutely incorrect results.

In oceanology and atmospheric sciences hydrostatic approximation is used to filter some 
classes of 
solutions of the full system like short gravity waves (buoyancy driven), convection, and sound waves~\cite{Charney1990,1979Kadyshnikov,1991ZeytounianMeteorologicalFluidDynamics}. It serves as an analogy to shallow water equations for surface waves and works at scales from hundreds of kilometers in Earth atmosphere. 
Of course, such a filtering can not be performed without sacrificing the quality of the forecast, and results in more complicated parameterizations, so the general current trend in modern weather prediction models is the transition to non-hydrostatic models~\cite{2020Gavrikov40yr}, improving parameterizations of turbulence and convective adjustment.

Smallness of the ratio of vertical force of inertia to the gravity is a \emph{property} of the quasihydrostatic approximation 
(like smallness of {\sl any} oil car velocity compared to the speed of light), 
and it can also be a property of certain non-hydrostatic flows.
But it can not be the reason for application of hydrostatic approximation as stated in the new theorem in~\cite{2018NigmatulinEquations},
since the major side-effect would be the wrong estimation of the scales, for which hydrostatic approximation can describe atmospheric (or oceanic) flows.
\vspace{-5mm}
\section{
  Incorrectness of scale analysis of 
  Richardson's equation 
  for vertical velocity%
.}
\vspace{-4mm}
L.F. Richardson in his fundamental work "Weather prediction by numerical process"  \cite{1922RichardsonWPNP} (1922) gave the framework for large-scale weather prediction based on hydrostatic approximation, with the expression for vertical velocity as an exact consequence of hydrostatic approximation in eq. 4 on p. 123, eq. 9 on p. 124, also see Peter Lynch "The Emergence of Numerical Weather Prediction: Richardson's Dream" \cite{2006LynchEmergence} eq. 2.19 on p. 40, A. Kasahara~(1967) \cite{1967KasaharaWashington1967} eq. 2.12, 2.13, A. Eliassen "Dynamic Meteorology" (1957) \cite{1957EliassenKleinschmidtDynamicMeteorology}, also his paper ~\cite{1949Eliassen} (1949), etc. 

In~\cite{2018NigmatulinEquations} (2018) the author took this expression without citation%
, and neglected the horizontal transport of pressure, 
upon comparing it
to \emph{one of the terms} of the horizontal mass divergence above the particle. 

This kind of comparison is not correct, since with the same reasoning the divergence of an incomprehensible fluid, \emph{being identically zero}, could be estimated to have a finite value. 
Scale of the horizontal divergence of mass flow also can differ by orders of magnitude from the scale of its components, especially in frames of hydrostatics, due to the following factors: 
1) principal reason of atmospheric compressibility (without acoustics) owes to the weight of atmospheric column and temperature changes, not to the high velocities; 
2) smallness of the full vertical acceleration%
; 
3) geostrophic component of the flows has zero divergence, 4) significant localized rise of pressure induce reaction in form of short gravity waves, which 
can not be described 
by quasistatic modeling
and are usually parametrized.
Mutual compensation of the components of horizontal divergence in large-scale motions of atmosphere was discussed in~\cite{Charney1990}.
The other issue is that the gradient of pressure in advection term is estimated in~\cite{2018NigmatulinEquations} \emph{dynamically} as $U^2/L$, so the horizontal velocity is interpreted here as the velocity of air in horizontal pressure waves (propagating with the speed of sound), and for such conditions the advection of pressure of course would be small. 
But if the pressure would be represented again \emph{hydrostatically} as the column weight, the estimation would be different.
Although advection of pressure can be neglected in some cases (as well as the horizontal divergence), for a general case, omitting the pressure advection
based on the scale analysis in~\cite{2018NigmatulinEquations} 
may {\sl violate the symmetries of the system} and allows a particle to accumulate additional uncompensated vertical velocity, which results in nonphysical effects.

From the above it follows, that 
reformulation of the L.F.~Richardsons's framework~\cite{1967KasaharaWashington1967,1922RichardsonWPNP} 
given 
in \cite{2018NigmatulinEquations} is not correct for modelling of ``climatic and meteorological processes'' and  weather forecast {\sl at any horizontal scales.} 

Richardson's framework was used at national center of atmospheric research NCAR~\cite{1967KasaharaWashington1967} (A.Kasahara et al., 1967) about a decade as the principal weather prediction system%
\footnote{
  With altitude as the vertical coordinate and Richardson's equation for vertical velocity. Nowadays, other variables are used as the vertical coordinate~\cite{1974KasaharaVariousCoordinates,2020Gavrikov40yr}, which are more advantageous for numerical modeling.
}%
, and a lot of works by several scientific groups are devoted to its stability and properties of the corresponding differential operator for different vertical discretizations and vertical coordinates~\cite{1974KasaharaVariousCoordinates,1978OligerSundstromIBVP,
2004BourchteinKadychnikov}.
\end{sloppypar}

\bibliography{quasiStatics}

\bibliographystyle{ieeetr}

\end{document}